\documentclass[preprint,aps,tightenlines,showkeys,nofootinbib,superscriptaddress]{revtex4}
\usepackage{latexsym,amssymb,amstext,amsmath}
\usepackage[dvips]{graphicx}
\usepackage[dvips]{color}
\newcommand{\beq}{\begin{eqnarray}}
\newcommand{\eeq}{\end{eqnarray}}
\newcommand{\bea}{\begin{eqnarray*}}
\newcommand{\eea}{\end{eqnarray*}}
\newcommand{\eq}{eqnarray}

\newcommand{\al}{{\alpha}}

\newcommand{\ga}{{\gamma}}

\newcommand{\de}{{\delta}}

\newcommand{\La}{{\Lambda}}
\newcommand{\m}{{\mu}}
\newcommand{\n}{{\nu}}
\newcommand{\si}{{\sigma}}

\newcommand{\ka}{{\kappa}}
\newcommand{\om}{{\omega}}

\newcommand{\pa}{{\partial}}
\newcommand{\no}{{\nonumber}}
\newcommand{\f}{\frac}
\newcommand{\ra}{\rightarrow}

\newcommand{\bq}{\bar{q}}
\newcommand{\vp}{\varphi}

\begin{document}

\preprint{arXiv:2101.1101v3 [hep-th]}
\title{No Scalar-Haired Cauchy Horizon Theorem in Einstein-Maxwell-Horndeski Theories}

\author{Deniz O. Devecio\u{g}lu \footnote{E-mail address: dodeve@gmail.com}}
\affiliation{School of Physics, Huazhong University of Science and Technology,
Wuhan, Hubei,  430074, China }

\author{Mu-In Park \footnote{E-mail address: muinpark@gmail.com, Corresponding author}}
\affiliation{ Center for Quantum Spacetime, Sogang University,
Seoul, 121-742, Korea }
\date{\today}

\begin{abstract}
{Recently, a no inner (Cauchy) horizon theorem for static black holes with
{\it non-trivial scalar hairs} has been proved in Einstein-Maxwell-scalar theories.
In this paper, we {study an extension of} the theorem to the static black holes in Einstein-Maxwell-Horndeski
theories. We study the black hole interior geometry for some exact solutions and find
that the spacetime has a (space-like) curvature singularity where the black hole mass gets
an extremum and the Hawking temperature vanishes. We discuss further extensions of the
theorem, including general Horndeski theories from
disformal transformations.}
\end{abstract}

\keywords{No hair theorem, Cauchy horizon, Black hole singularity, Horndeski theory}

\maketitle

\newpage
Exploring the interior structure of black holes has been a fundamental topic in general relativity (GR)
but much are not well understood yet. For example, whether the interior spacetime is singular or {not}
for the real black holes in nature as observed in various ways \cite{Abbo:2016,Akiy:2019}. The other
question, which is relevant to this paper is whether the (classical) predictability in GR is lost
due to the inner (Cauchy) horizons as in the exact solutions, like Reissner-Nordstrom, Kerr black holes.
Regarding the latter question, recent development suggests that Cauchy horizons for some kind of
charged black holes are unstable in the presence  of scalar hairs \cite{Hart:2012,Cai:2020}. In
particular, in \cite{Cai:2020} it is proved that there exists no inner Cauchy horizons for both
planar and spherical static black holes with {\it non-trivial scalar hairs} in Einstein-Maxwell-scalar
theories. From the simplicity and quite generic results, which do not  depend on the details of the
scalar potentials as well as the spacetime structure at the short distance, it's applicability would be
quite far-reaching. In this respect, it would be important to classify all possible extensions of
the theorem.

In this paper, as the first step towards that goal, we consider an extension of the theorem to
the black holes in Einstein-Maxwell-Horndeski (EMH) theories, with
{a quartic Horndeski term}
\cite{Hord:1974}. To this ends, we consider the $D$-dimensional EMH action, with the $U(1)$ gauge filed $A_{\mu}$ and {\it charged} scalar field $\varphi$,
\begin{\eq}
I&=&\int d^D x \sqrt{-g} \left[ \f{1}{\kappa} R +{\cal L}_m \right], \label{action}\\
{\cal L}_m &=& -\f{Z(|\varphi|^2)}{4 \kappa} F_{\mu \nu} F^{\mu \nu}-\left( \al g^{\mu \nu} -\gamma G^{\mu \nu}\right) (D_{\mu} \varphi)^* (D_{\nu} \varphi)-V(|\varphi|^2), \no
\end{\eq}
where $F_{\mu \nu} \equiv \nabla_{\mu} A_{\nu}-\nabla_{\nu} A_{\mu}$ and $ D_{\mu} \equiv \nabla_{\mu}-i q A_{\mu}$ with a scalar field charge $q$ and $\kappa \equiv 16 \pi G$. Here, we have introduced $Z$ and $V$ as arbitrary functions of $|\varphi|^2$ in which the constant shift symmetry of the Horndeski terms, with { a non-minimal (derivative) coupling $\ga$ of the scalar field to the Einstein tensor $G^{\mu \nu} \equiv R^{\mu \nu}-(1/2) g^{\mu \nu} R$} as well as the usual minimal coupling { $\al$}, may be broken generally. { Note that this generalizes the recent set-up in \cite{Hart:2012,Cai:2020} into {\it higher-derivative} theories via the Einstein coupling.}

Let us then consider the general static ansatz,
\begin{\eq}
ds^2&=&-N^2(r)dt^2+\f{1}{f(r)}dr^2+r^2 d \Omega^2_{D-2,k},
\label{Ansatz}\\
 \varphi&=&\varphi(r),~A=A_t(r)dt, \no
\end{\eq}
where $d \Omega^2_{D-2,k}$ is the metric of $(D-2)$-dimensional unit sphere with the spatial curvature which is normalized as $k=1,0,-1$ for the spherical, planar, hyperbolic topologies. The equations of motion [Here, we consider $D=4$ case for simplicity. But its higher-dimensional generalization is straightforward  (for the details, see Appendix {\bf A}) \cite{Cai:2020,Feng:2015} and we will discuss about this later] are given by (the prime $(')$ {and the dot $(\dot{~})$ denote the derivatives with respect to $r$ and $|\varphi|^2$, respectively}),
\begin{\eq}
E_{A}&\equiv& \f{1}{\ka} \left( Z(|\varphi|^2) \f{\sqrt{f}}{N}~ r^{2} A_t' \right)'
-\f{2 q^2 |\varphi|^2 A_t }{f} \left(\f{\sqrt{f}}{N} \right) \left[\al r^2+\ga (k-f)-\ga r f' \right]=0,
\label{E_A}\\
E_{\vp}&\equiv& \left(\f{N}{\sqrt{f}} f \left[\al r^2+\ga (k-f)-\ga r f \left( \f{{N^2}'}{N^2}\right) \right] \vp' \right)' \no \\
&&+\left(\f{N}{\sqrt{f}} \right)   \f{ q^2  A_t^2 \varphi  }{N^2} \left[\al r^2+\ga (k-f)-\ga r f' \right] +r^{2} \left[\f{1}{2 \ka} \f{\sqrt{f}}{N} \dot{Z} {A_t'}^2-\f{N}{\sqrt{f}}~ \dot{V} \right] \vp=0, \label{E_varphi}\\
E_{N} &\equiv & \f{2}{\ka} \left(f-k+{r f'}\right)+r^2 V (|\varphi|^2)
+f \left[\al r^2+\ga (k+f) +3  \ga r f' \right] |\varphi'|^2  \no \\
&&+2 \ga r f^2 \left( |\varphi'|^2\right)' +\f{1}{2 N^2}\left( \f{1}{\ka} {Z}f r^2 {A_t'}^2
+2 q^2 |\varphi|^2 A_t^2 \left[\al r^2+\ga (k-f)
-\ga r f' \right] \right)=0,
\label{E_N}\\
E_f &\equiv& \f{2}{\ka} \left[ f-k+{r f} \left( \f{{N^2}'}{N^2}\right) \right]+r^2 V (|\varphi|^2) -f\left[\al r^2+\ga (k-3f)-3  \ga r f \left( \f{{N^2}'}{N^2}\right) \right] |\varphi'|^2 \no \\
&&+\f{1}{2 N^2}\left( \f{1}{\ka} {Z}f r^2 {A_t'}^2
-2 q^2 |\varphi|^2 A_t^2 \left[\al r^2+\ga (k-f)\right] \right) \no \\
&&+\f{\ga q^2}{N^2}\left(|\varphi|^2 \left[-r f A_t^2 \left( \f{{N^2}'}{N^2}\right)+ 2 r f (A_t^2)'  \right] +2 r f A_t^2 (|\varphi|^2)'  \right)=0,
\label{E_f}
\end{\eq}
where the first two equations (\ref{E_A}), (\ref{E_varphi}) are the equations for the gauge field and scalar field respectively and the last two (\ref{E_N}), (\ref{E_f}) are for the metric. In particular, (\ref{E_A}) and (\ref{E_N}) correspond to the {\it Gauss's law constraint} and
{\it Hamiltonian constraint} equations in canonical formulation, respectively.\footnote{{ Here, $N$ denotes the lapse function in the standard ADM decomposition. Note that $N$ and $f$ appear together as ``$N/\sqrt{f}$ " in $\sqrt{-g}$ whereas they appear independently as $N^2$ and $f$ in the Lagrangian density ${\cal L}=(1/\ka) R+{\cal L}_m$. So, the action (\ref{action}) and its resulting all equations of motion (\ref{E_A}-\ref{E_f}) are real even inside the event horizon where $N^2, f<0$ so that $N$ and $\sqrt{f}$ become imaginary.}} Even though it is difficult to solve the coupled non-linear PDE generally, several exact solutions are known for some specific choices of couplings and potentials. We will study the black hole interior geometry for those exact solutions in details. But before this, we first prove the ``no scalar-haired Cauchy horizon theorem" quite generally, without knowing the details of the interior geometry,
{which extends the argument of
\cite{Cai:2020} for Einstein-Maxwell-scalar theories into the EMH theory with a non-vanishing Einstein coupling $\ga$}.\\

{\it No Scalar-Haired Cauchy Horizon Theorem}: The proof depends on whether the scalar field is
charged or not. Let us first consider the charged scalar field case \footnote{We may take a {\it real}
scalar field because the phase field $\om$ of the ``complex" scalar field with a charge
$\vp=|\vp| e^{i \om}$ can be consistently fixed to be zero in the equations of motion (\ref{E_A})
-~(\ref{E_f}), without affecting the coupling to gauge field. We may further consider
$f(r) \equiv N^2 e^{W(r)}$ with a {\it smooth} (finite) function $W(r)$ at the horizons,
for the convenience of the proof without loss of generality \cite{Cai:2020}. The holography-inspired metric in the literature \cite{Hart:2012,Cai:2020} can be also obtained from the coordinate transformation, $r=1/z, N^2(r)=e^{-\chi(z)} \tilde{f} (z)/z^2 , f(r)=\tilde{f}(z)/z^2$. }. Then, suppose that there were a {\it finite-temperature}
({\it i.e.}, non-extremal) black hole with the outer and inner (Cauchy) horizons at $r_+$ and $r_-(<r_+)$, respectively (Fig. 1), {\it i.e.}
\begin{\eq}
&&N^2 (r_+)=f (r_+)=0,~{N^2}'(r_+), f'(r_+)>0, \label{BC_rH}\\
&&N^2 (r_-)=f(r_-)=0,~{N^2}'(r_-), f'(r_-) \leq 0. \label{BC_rC}
\end{\eq}
Now, in order to have a ``non-trivial", {\it i.e.}, finite {\it charged} scalar field $\vp$ at the two
horizons, the equations of motion (\ref{E_A}
-~\ref{E_f}) imply the conditions \footnote{This phenomena which represents a ``repulsion between the charged scalar field and the gauge field", similar to {\it Meissner effect} in superconductivity. This seems to be quite generic \cite{Cai:2020} and might reflect {\it the black hole horizon superconductivity}. } \cite{Cai:2020} {(for a more rigorous derivation, see Appendix {\bf B})},
\begin{\eq}
A_t (r_+)=A_t (r_-)=0
\label{BC_A}
\end{\eq}
{from the regularity at the
horizons. This regularity condition is one of the key ingredients in the proof. }

\begin{figure}
\includegraphics[width=10cm,keepaspectratio]{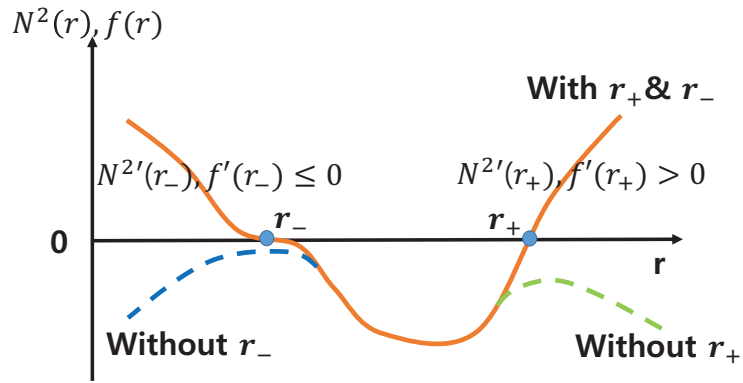}
\caption{Typical plots of $N^2(r)$ and $f(r)$ for a {\it finite-temperature}, charged black hole with the inner (Cauchy) horizon $r_-$ and outer (event) horizon $r_+$. Due to a non-zero Hawking temperature, we have $N^2=f=0,~{N^2}',f'>0$ at the outer horizon $r_+$, whereas $N^2= f=0,~{N^2}', f'\leq 0$ at the inner horizon generally, {\it if existed} (orange, solid line). In the absence of an inner horizon, ${N^2}', f'$ can have arbitrary values but there is no inner point of $N^2=f=0$ (blue, dashed line). On the other hand, in the absence of outer horizon but now with an inner horizon, the situation is similar for an observer sitting inside the inner horizon which becomes now an event horizon  (green, dashed line).}
\label{fig:N_f_AdS}
\end{figure}

Another key ingredient is the existence of {\it radially} conserved charge,
\begin{\eq}
{\cal Q}&=& (D-2) r^{D-2} \left[N \sqrt{f} \left(\f{1}{\ka} +\f{\ga}{2} f |\varphi'|^2 \right) \left(\f{{N^2}'}{N^2}-\f{2}{r} \right)-\f{1}{2 \ka} \f{\sqrt{f}}{N} Z  {A_t^2}' \right. \no  \\
 &&+\left.  \f{\sqrt{f}}{N} \ga q^2 A_t^2  |\varphi|^2 \left( \f{ 2 A_t'}{A_t}-\f{{N^2}'}{2 N^2}-\f{1}{r} \right)+  \f{\sqrt{f}}{N} \ga q^2 A_t^2  {|\varphi|^2}' \right] \no \\
&&+k (D-2) (D-3) \int ^r dx~ x^{D-4} \left[\f{\ga q^2 A_t^2  |\varphi|^2 }{ \sqrt{N^2 f}}+ \f{N}{\sqrt{f}} \left(\f{2}{\ka} -{\ga} f |\varphi'|^2 \right) \right],
\label{Q}
\end{\eq}
where we have recovered the dimensional dependencies. From
the equations of motion (\ref{E_A})
-~(\ref{E_f}), one can prove that ${\cal Q}$ is radially conserved, {\it i.e., r-independent},
due to a (remarkable) relation \footnote{In the black hole interior, where $r$ is a time
coordinate, this becomes a {\it dynamical} relation whose detailed physical meaning is still
unclear and needs to be investigated.},
\begin{\eq}
{\cal Q}'&=&2 \f{\sqrt{f}}{N} \left[ r \left( \f{N^2}{f} {{\cal E}_f}'+\f{1}{2} \f{{N^2}'}{f} {\cal E}_f \right)
+\left( (D-2) \f{N^2}{f} +\f{r}{2} \f{{N^2}'}{f}\right) {\cal E}_N \right. \no \\
&&\left.+\f{r}{2} N^2 \vp'~ {\cal E}_\vp+\f{1}{2} \left( (D-2) A_t + r A_t' \right) {\cal E}_A \right] =0,
\end{\eq}
where ${\cal E}_{\psi}$'s represent the equations of motion for the fields $\psi=(N,f,\vp,A_t)$ which
can be obtained directly from the action and related to $E_{\psi}$'s in (\ref{E_A})
-(\ref{E_f}) as ${\cal E}_N =-E_N r^{D-4}/2,~ {\cal E}_f =E_f r^{D-4}/2,~ {\cal E}_\vp =2 E_\vp/N \sqrt{f}, ~ {\cal E}_{A} =- E_{A}N/\sqrt{f}$.

For the planar topology, $k=0$, the first line of (\ref{Q}) is a Noether charge associated with a particular scaling symmetry for a neutral scalar field with $q=0$ \cite{Feng:2015,Gubs:2009}. For other topologies $k \neq 0$, the {\it local} parts in the first and second lines are not conserved but the {\it non-local}, integrated terms are needed in order that the non-conserved terms are canceled \cite{Cai:2020}.

Now, due to the $r$-independence, the charge satisfies ${\cal Q}_{r_+}={\cal Q}_{r_-}$ at the two horizons, in particular, where the boundary conditions (\ref{BC_rH}), (\ref{BC_rC}), and (\ref{BC_A}) hold, we have the relation \footnote{For $D=3$, the result is essentially the same as in the planar topology due to the absence of spatial curvature so that the theorem applies always (we thank the authors of \cite{Cai:2020} for the discussion about this matter). For $D=2$, the charge ${\cal Q}$ of (\ref{Q}) vanishes trivially so that there is no constraint on $D=2$ black holes from (\ref{Q_rel}). This is due to a triviality of the $D=2$ Einstein gravity and a separate consideration is needed for an appropriate treatment. But, since the proper $D=2$ gravity, which being a {\it dilaton gravity}, can be obtained from a dimensional reduction of $D=3$ gravity, the result is not much different from that of $D=3$ for the conventional Kaluza-Klein reduction (for an explicit computation, see Appendix {\bf C}). }
\begin{\eq}
&&\f{(D-2)}{\ka} \left[ r_+^{D-2} \f{\sqrt{f}}{N} \left.{N^2}' \right|_{r_+}
-r_-^{D-2} \f{\sqrt{f}}{N} \left.{N^2}'\right|_{r_-} \right] \no \\
&=&k (D-2) (D-3) \int ^{r_+}_{r_-} dx~ x^{D-4} \left[-\f{\ga q^2 A_t^2  |\varphi|^2 }{ \sqrt{N^2 f}}+ \f{N}{\sqrt{f}} \left(-\f{2}{\ka} +{\ga} f |\varphi'|^2 \right) \right].
\label{Q_rel}
\end{\eq}
From the assumed inner horizon $r_-$, {\it i.e.}, $ {N^2}'\sqrt{f}/N |_{r_-} \leq 0$, as well as the
outer horizon $r_+$ with a non-vanishing Hawking temperature or surface gravity
$\sim  {N^2}' \sqrt{f}/N|_{r_+} >0$, the left hand side of (\ref{Q_rel}) is {\it positive}. On the other hand,
the integrand on the right hand side is {\it negative} for the ``positive" Horndeski coupling
constant $\ga>0$ due to the space-time structure between $r_+$ and $r_-$, where $ N^2, f<0$, so that
(\ref{Q_rel}) is inconsistent unless the hyperbolic topology $k=-1$ is considered. In other words,
{\it no smooth inner (Cauchy) horizon with the outer horizon} can be formed for $k=0$ or $k=1$ and
this generalizes the theorem of \cite{Hart:2012, Cai:2020} to {a class of the EMH theories (\ref{action}) with the positive Einstein coupling $\ga>0$. However, for a negative Einstein coupling $\ga<0$, one can not get a strong constraint on the inner horizon, due to lack of a definite sign of the integrand on (\ref{Q_rel}), except the case $k=0$ in which the above theorem still works.\footnote{It seems that the sign of $\ga$ is correlated with the speed of gravitational waves $c_{GW}$ for the black hole background (\ref{Ansatz}). In particular, for the $D=4$ asymptotically flat, charge-neutral spherical black holes with a real scalar field ($\al>0$), $c^2_{GW}=\f{2 \ga +\al r^2}{2\ga+3\al r^2}\leq 1$ for $\ga>0$ \cite{Koba:2014}.
}} Here, it interesting to note that the theorem may also imply the {\it no outer horizon with an inner horizon} as described in Fig.1 (green, dashed line). This completes the
proof of the theorem for the {\it charged} scalar field with $q \neq0$ and note that the proof
is quite generic without assuming any specific form of the scalar potential.

On the other hand, for
the $k=-1$ {($\ga>0$)} case, the relation (\ref{Q_rel}) {\it can} be consistent with the inner horizon but it can be still consistent even {\it without} the inner horizon, {\it i.e.},
$ {N^2}'\sqrt{f}/N |_{r_-} > 0$ unless its contribution dominates that of the outer horizon
in the left hand side so that (\ref{Q_rel}) is still satisfied. In this sense, the theorem does not
give any strong constraint about the inner horizon formation for the $k=-1$ case, except the possibility
of its formation. Moreover, for the neutral scalar field ($q=0$), the vanishing gauge field (\ref{BC_A}) can not be
justified anymore and there is no constraint on gauge field from the regularity at the
horizons
so that the novel relation (\ref{Q_rel}) can not be used for the proof and we need a separate consideration as described below.
However, we note that the same theorem also applies to {\it neutral} black holes, {\it i.e.},
in the absence of $U(1)$ gauge field $A_{\mu}$, since the essential parts of (\ref{Q_rel})
do not depend on it.

For the proof of {\it neutral} scalar field, following \cite{Hart:2012}, we multiply $\vp^*$ to (\ref{E_varphi}) and obtain
\begin{\eq}
 0&=&\int^{r_+}_{r_-} \left(\f{N}{\sqrt{f}}~ f \left[\al r^2+\ga (k-f)-\ga r f \left( \f{{N^2}'}{N^2}\right) \right] \vp^* \vp' \right)' \no \\
&=&\int^{r_+}_{r_-} \f{N}{\sqrt{f}}
\left\{ f \left[\al r^2+\ga (k-f)-\ga r f \left( \f{{N^2}'}{N^2}\right) \right] |\vp'|^2
- r^2 \left[\f{1}{\ka} \f{f}{N^2} \dot{Z}  {A_t'}^2 -\dot{V}\right] |\vp|^2 \right. \no \\
&&~~~~~~~~~~\left. -  \f{ q^2 A_t^2 |\varphi|^2  }{N^2} \left[\al r^2+\ga (k-f)-\ga r f' \right]
 \right\}. \label{E_varphi_int}
\end{\eq}
For the {\it neutral} scalar field with $q=0$, where the last term in the third line is absent, the integrand in the second line is {\it non-positive} if the quantities in $[~~]$ for first and second terms in the second line are ``positive". In this case, the second line is consistent with the first line only for the vanishing scalar field $\vp=0$ in the range of $(r_-, r_+)$, which means the impossibility of inner (Cauchy) horizon with a non-trivial scalar field ({\it i.e., hair}) and an outer (event) horizon, or similarly the impossibility of an outer horizon with a non-trivial scalar field and an inner horizon, as in the charged case. For the Einstein-Maxwell-scalar theories with $\al>0, \ga=0$ as in \cite{Hart:2012,Cai:2020}, the required condition for the theorem is $ \dot{Z}  {A_t'}^2 {f}/{N^2} \ka -\dot{V}>0$. With the Horndeski
{coupling
$\ga$},
we may need a
further condition, like $\al r^2+\ga (k-f)-\ga r f  {{N^2}'}/{N^2}>0$. This
produces the constraint on the range, $r<
{{(N^2/f)} [\al r^2+\ga (k-f)]}/{{\ga}{N^2}'}$, near $r_+$ (or equivalently near $r_-$) with ${\ga} {N^2}'>0$ for the validity of the theorem, which {\it could} imply the non-vanishing scalar field in the black hole interior near $r_+$ {(or exterior near $r_-$)}. On the other hand, in the presence of a charge $q$, the last term gets a positive contribution even if the above conditions are satisfied so that one can not generally rule out the inner (or outer) horizon formations, as noted in \cite{Cai:2020}, unless we need more conditions on the metric and matter fields or electric charge.

{\it An Example of Interior Geometry with Inner Horizon}

Now, in order to study explicitly the interior geometry of a charged black hole with a non-trivial scalar field, we consider an exact solution for $\al=0$ (no minimal scalar coupling), $q=0$ (neutral scalar field), $Z=1$ (minimal gauge coupling), $V=2 \La/\kappa$ (cosmological constant term, no self-scalar-coupling) in $D=4$, for simplicity (for $k=1$, see \cite{Anab:2014,Cist:2014,Feng:2015}\footnote{For the $k$-dependent terms, there is a $k$-factor difference with \cite{Anab:2014}. This gives the different results for $k=-1$. On the other hand, for $k=0$, where the solution (\ref{vp_sol}) is not defined, we need a different solution \cite{Anab:2014}.} and also \cite{Rina:2012,Babi:2013}) but $\ga \neq 0$ (non-vanishing Einstein tensor-scalar coupling),
\begin{\eq}
N^2&=&k \left(k^2+\f{\La \bar{q}^2}{8} \right)-\f{M}{r} +\f{k^2}{4} \f{\bar{q}^2}{r^2}-\f{k}{192} \f{\bar{q}^4}{r^4}-\f{k^2\La}{3}r^2 +\f{k \La^2}{20} r^4,
\label{f_sol}\\
\f{f}{N^2} &=&\f{64 r^4}{(4 \La r^4 -8 k  r^2 +\bar{q}^2)^2},
\label{N_sol} \\
A_t &=&A_t^{(0)}-\f{k \bar{q}}{r}+\f{\bar{q}^3}{24 r^3}-\f{\bar{q} \La}{2} r,
\label{A_sol}\\
\vp' &=& \sqrt{\f{4 \La r^2+{\bar{q}^2}/{r^2}}{-4 \ga k \ka f}},
\label{vp_sol}
\end{\eq}
where $M$ and $\bar{q}$ represent the two integration constants, associated with the black hole mass and charge, respectively. Here, we note that the solution can be defined only when $\ga \neq 0$ is considered though not manifest in the metric and gauge field solutions. But interestingly, the solution for $k=\pm1$, pure Einstein-Maxwell theories with a scalar field, {\it i.e.}, $\al=\ga=0$, can be also read with a certain truncation of higher-order terms,
\begin{\eq}
N^2&=&f=k -\f{M}{r} +\f{1}{4} \f{\bar{q}^2}{r^2}-\f{\La}{3}r^2 ,
\label{f_sol_GR}\\
A_t &=&A_t^{(0)}-\f{k \bar{q}}{r},
\label{A_sol_GR}
\end{\eq}
which seems to imply a smooth limit of $\ga \ra 0$.
\begin{figure}
\includegraphics[width=10cm,keepaspectratio]{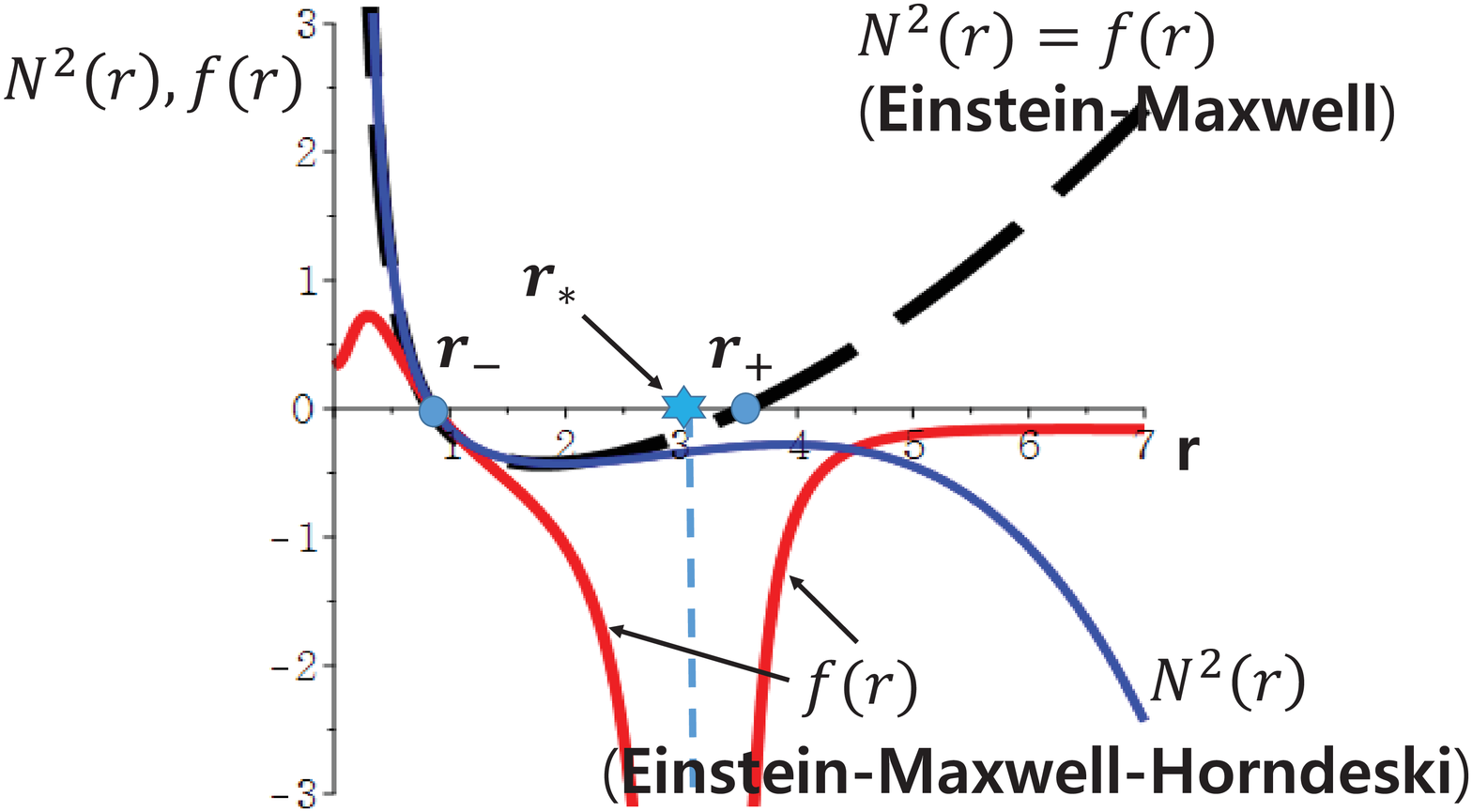}
\caption{Plots of $N^2(r)$ (blue line) and $f(r)$ (red line) for a charged black hole solution (\ref{f_sol}), (\ref{N_sol}) with $k=-1, M<0$ in ant-de Sitter space $\La<0$. Due to a Horndeski coupling $\ga\neq 0$, the metric function $f$ is divergent at $r_* \approx 3.18$ where a (space-like) curvature singularity occurs, in addition to the usual singularity at $r=0$. Moreover, even with the inner horizon $r_-$, no outer horizon $r_+$ is formed (red, blue solid lines), in contrast with two horizons in the pure Einstein-Maxwell theories (dashed line). Here, we have considered $\La=-0.2, q=1, M=-0.5$.}
\label{fig:N_f_AdS}
\end{figure}

The metric
has a surface-like curvature singularity (we omit the explicit expressions of curvatures due to their messy expressions) at
\begin{\eq}
r_*=\sqrt{\f{-k \mp \sqrt{k^2-\La \bar{q}^2/4}}{-\La}}
\end{\eq}
in which $f/N^2$ in (\ref{N_sol}) diverges, as well as the usual point-like singularity at $r=0$ (see Fig. 2, for the case $k=-1, M<0$). When the black hole horizon $r_H$ reaches to the singular radius $r_*$, {\it i.e.} $r_H=r_*$, the black hole mass parameter $M$,
\begin{\eq}
M=k \left(k^2+\f{\La \bar{q}^2}{8} \right) r_H +\f{k^2}{4} \f{\bar{q}^2}{r_H}-\f{k}{192} \f{\bar{q}^4}{r_H^3}-\f{k^2 \La}{3}r_H^3 +\f{k \La^2}{20} r_H^5
\label{M}
\end{\eq}
becomes an extremum value,
\begin{\eq}
M_*=\f{[k (\La^2 \bq^4 -12 k^2 \La \bq^2-32 k^4) \mp 16 (2k^2 +\La \bq^2) \sqrt{k^2-\La \bq/4}]}
{30 \sqrt{\La} (k \pm \sqrt{k^2-\La \bar{q}^2/4})^{3/2}}.
\label{M_*}
\end{\eq}
Moreover, at the extreme horizon radius $r_H=r_*$, the Hawking temperature \footnote{In this convention of the Hawking temperature $T_H$, we have $T_H>0$ for an observer at $r<r_-$ and $k=-1$, as well as an observer at $r>r_+$ and $k=+1$. This may be also compared with the Hawking temperature for the pure Einstein-Maxwell case, $ T_H=\f{\hbar k}{32 \pi r_H^3} \left(  8 k^2 r_H^2 -2\bq^2-8  \La r_H^4\right)$.},
\begin{\eq}
T_H&=&\left.\f{\hbar}{4 \pi} \left( \f{N}{\sqrt{f}}\right) {N^2}' \right|_{r_H} \no \\
   &=& \f{\hbar k}{32 \pi r_H^3} \left(  8 k^2 r_H^2 -k\bq^2-4 k \La r_H^4\right)
\end{\eq}
becomes zero.

For the hyperbolic case with $k=-1$ and $\La, M<0$, there is only a ``cosmological" horizon $r_-$, which {\it used to} be the inner (Cauchy) horizon in the purely Einstein-Maxwell theories without a (neutral) scalar field (dashed line in Fig. 2). In this case, the outer horizon is not formed due to Horndeski coupling $\ga$ of scalar field \footnote{For the spherical case with $k=+1$ and $M>0,~ \La<0$, the inner horizon is not formed instead and there is only a black hole horizon $r_+$.}, which behaves as (with a vanishing integration constant)
\begin{\eq}
&\vp &=2 \left.\sqrt{\f{r g(r)}{2 \ga \ka}}\right|_{r_-}~ \sqrt{r_- -r} +{\cal O} (r_- -r)^{3/2}, \\
&g(r) &\equiv 4 \La r^2+{\bar{q}^2}/r^2,
\end{\eq}
near the inner horizon $r_-$, implying the vanishing and non-analytic \footnote{From the non-analyticity, $\vp'$ in (\ref{vp_sol}) is singular at the horizon $r_H$, where $f(r_H)=0$. But a coordinate-invariant quantity $g^{\mu \nu}\pa_\mu \vp \pa_\nu \vp=f(r) {\vp'}^2$ is regular for the whole region, including the singular point $r_*$.} scalar field at the inner horizon when $g(r_-)/\ga>0$. In the exterior region of the (inner) horizon, $r>r_-$, where $r$ is a time coordinate, the scalar field may also exist and behaves as, though non-analytic again, near the surface-like singularity $r_*$,
\begin{\eq}
\vp =-\f{4}{3} \left. \sqrt{\f{4 k^2-k \bq^2/r^2}{2  \ga k \ka N^2}}~\right|_{r_*} (r_* -r)^{3/2} +{\cal O} (r_* -r)^{5/2}
\end{\eq}
when $g(r)/\ga<0$, {\it i.e.}, $r_- < \tilde{r} < r$ with $g(\tilde{r})=0$ and $\varphi' (r_*)=0$.
On the other hand, the gauge field is regular and analytic at the (inner) horizon and has  a vanishing electric field, $A_t'=\f{\bq}{r^{2}}\f{N}{\sqrt{f}}$ at $r_*$.


{{\it Concluding Remarks}}

As a summary, we have an extension of the ``no scalar-haired inner (Cauchy) horizon theorem"
to the black holes in {a class of the EMH theories with a
Einstein coupling $\ga$, which crucially depends on the sign of $\ga$:
No {\it charged} scalar-haired Cauchy horizon in $k=0,1$ (planar, spherical) black
holes for $\ga>0$, whereas only in $k=0$ black holes for $\ga<0$. Here, it is important to note
that the proof is quite generic without assuming any specific form of the scalar potential nor the spacetime structure at the short distance. On the other hand, for the {\it neutral} scalar field, we need some {\it kinematic} conditions for the corresponding theorem, which depends on the details of the system.}
As an explicit example of the black hole interior geometry with inner horizon, we studied some exact solutions with an inner horizon but no outer horizon due to non-trivial scalar field (hairs). Several further remarks about remaining challenging problems are in order.

1. We have considered a special case of Horndeski terms, with {$G_2$ and $G_4$} in the usual terminology. By considering {\it disformal transformations}, $\tilde{g}_{\mu \nu}={A(\varphi)} g_{\mu \nu}+{B(\varphi)} \nabla_{\mu} \varphi \nabla_{\nu} \varphi$, with $X \equiv -g^{\mu \nu} \nabla_{\mu}  \nabla_{\nu} \varphi$ one can get the general form of the Horndeski action but with the special coefficients depending on $A$ and $B$ \cite{Bett:2013}. Hence, for the obtained Horndeski actions, one can also argue the same no-Cauchy-horizon theorem. It would be interesting to see whether we have the same result for the {\it fully general} Horndeski actions or not.

2. The theorem implies that the instability of Cauchy horizon under non-trivial scalar field (linear or non-linear) perturbations. However, the (dynamical) stability of the non-trivial hairy geometry is not be guaranteed generally and its explicit check would be an important complementary for the theorem.

3. In this paper, we have considered the Cauchy horizon for a ``spherically symmetric", charged black hole. It would be a challenging question whether the theorem can be generalized to the {\it non-spherically} symmetric case also, if the removal of Cauchy horizon by the non-trivial scalar hairs is a real one. One outstanding case would be {\it rotating} black holes and it would be interesting to see whether an angular velocity $\Omega$ and {\it spinning} matter fields ``repel" each other ``at the horizon", like the gauge field and {\it charged} scalar fields in the charged black holes (see footnote No.2), due to similarity of angular velocity and gauge connection \cite{Yoon:0005}.


\section*{Acknowledgments}
We would like to thank Rong-Gen Cai, Li Li, Tsutomu Kobayashi, and Kyung Kiu Kim for helpful discussions. DOD was supported by the National Natural Science Foundation of China under Grant No. 11875136 and the Major Program of the National Natural Science Foundation of China under Grant No. 11690021. MIP was supported by Basic Science Research Program through the National Research Foundation of Korea (NRF) funded by the Ministry of Education, Science and Technology {(2016R1A2B401304, 2020R1A2C1010372, 2020R1A6A1A03047877)}.

\appendix{\label{app1}}
\begin{section}
{Full Dimension-Dependent Equations of Motion}
\end{section}

In this Appendix, we present the full dimension-dependent equations of motion for  $D=4$ case (\ref{E_A})
-~(\ref{E_f}) in the text:

\begin{\eq}
E_{A}&\equiv& \f{1}{\ka} \left( Z(|\varphi|^2) \f{\sqrt{f}}{N}~ r^{D-2} A_t' \right)' \no \\
&&-\f{2 q^2 r^{D-4} |\varphi|^2 A_t }{f} \left(\f{\sqrt{f}}{N} \right) \left[\al r^2+\f{(D-2)(D-3)}{2}\ga (k-f)-\f{(D-2)}{2}\ga r f' \right]=0,
\label{E_A_Full}\\
E_{\vp}&\equiv& \left(\f{N}{\sqrt{f}}r^{D-4} f \left[\al r^2+\f{(D-2)(D-3)}{2}\ga (k-f)-\f{(D-2)}{2}\ga r f \left( \f{{N^2}'}{N^2}\right) \right] \vp' \right)' \no \\
&&+\left(\f{N}{\sqrt{f}} \right)   \f{ q^2 r^{D-4} A_t^2 \varphi  }{N^2} \left[\al r^2+\f{(D-2)(D-3)}{2}\ga (k-f)-\f{(D-2)}{2}\ga r f' \right] \no \\
&&+r^{D-2} \left[\f{1}{2 \ka} \f{\sqrt{f}}{N} \dot{Z} {A_t'}^2-\f{N}{\sqrt{f}}~ \dot{V} \right] \vp=0, \label{E_varphi_Full}\\
E_{N} &\equiv & \f{(D-2)(D-3)}{\ka} \left(f-k+\f{r f'}{D-3}\right)+r^2 V (|\varphi|^2) \no \\
&&+f \left[\al r^2+\f{(D-2)(D-3)}{2}\ga (k+f) + \f{3(D-2)}{2} \ga r f' \right] |\varphi'|^2 +(D-2) \ga r f^2 \left( |\varphi'|^2\right)' \no \\
&&+\f{1}{2 N^2}\left( \f{1}{\ka} {Z}f r^2 {A_t'}^2
+2 q^2 |\varphi|^2 A_t^2 \left[\al r^2+\f{(D-2)(D-3)}{2}\ga (k-f)
-\f{(D-2)}{2}\ga r f' \right] \right)=0, \no \\
\label{E_N_Full}\\
E_f &\equiv& \f{(D-2)(D-3)}{\ka} \left[ f-k+\f{r f}{D-3} \left( \f{{N^2}'}{N^2}\right) \right]+r^2 V (|\varphi|^2) \no \\
&&-f\left[\al r^2+\f{(D-2)(D-3)}{2}\ga (k-3f)- \f{3(D-2)}{2} \ga r f \left( \f{{N^2}'}{N^2}\right) \right] |\varphi'|^2 \no \\
&&+\f{1}{2 N^2}\left( \f{1}{\ka} {Z}f r^2 {A_t'}^2
-2 q^2 |\varphi|^2 A_t^2 \left[\al r^2+\f{(D-2)(D-3)}{2}\ga (k-f)\right] \right) \no \\
&&+\f{(D-2)}{2} \f{\ga q^2}{N^2}\left(|\varphi|^2 \left[- r f A_t^2  \left( \f{{N^2}'}{N^2}\right)+ 2 r f (A_t^2)'  \right] +2 r f A_t^2 (|\varphi|^2)'  \right)=0.
\label{E_f_Full}
\end{\eq}

\section{Proof of $A_t=0$ at the Horizons for Hairy Charged Black Holes}
{\label{appNew}}

In this Appendix, we present a more rigorous proof of the condition
(\ref{BC_A}), $A_t=0$ at the horizons $r_H$, for hairy charged black holes,
which is one of the key ingredient in the ``no scalar-haired Cauchy horizon theorem".

To this end, we first note that the equations of motion (\ref{E_A}
-~\ref{E_f}) become {\it singular} unless `$q \varphi A_t$=0' at the horizons.  [Here, we consider $D=4$ case for simplicity, but the conclusion is unchanged for arbitrary higher dimensions] In order to avoid the singular terms for the {\it charged} scalar field, {\it i.e.}, $q \neq 0$, we can consider (a) $A_t|_{r_H}=0$, or (b) $A_t|_{r_H} \neq 0$, $\varphi|_{r_H}=0$. Then, for the {\it hairy} black holes, {\it i.e.}, $\varphi$ is {\it smooth} and non-vanishing somewhere, we can prove that the case (a) is the only possible choice.

The proof is a {\it proof by contradiction}, following \cite{Cai:2020}: Let us first suppose that the case (b) is true. Then, since $\varphi$ is smooth and non-vanishing somewhere, one can consider the Taylor expansion near the horizons $r_H$,
\begin{\eq}
\varphi(r)=\varphi_l \de^l + \varphi_{l+1} \de^{l+1} +\cdots
\end{\eq}
with $\varphi_l \equiv \f{1}{l !} \varphi^{(l)}|_{r_H} \neq 0$ ($l \geq 1$) and $\de\equiv r-r_H$. Similarly, one can consider the expansions for the metric functions,
\begin{\eq}
N^2(r)&=&N^2_{n}\de^n + N^2_{n+1} \de^{n+1} +\cdots, \\
f(r)&=&f_{n}\de^n +f_{n+1} \de^{n+1} +\cdots
\end{\eq}
with $N^2_{n}, f_{n} \neq 0$ ($n \geq 1$), from the smoothness of $f(r)/N(r)^2\equiv e^{W(r)}$ (see foonote No. 2). From the smoothness of (\ref{E_varphi}), we need $l \geq n$, but let us consider $n=1$ for simplicity. Then (\ref{E_varphi}) becomes
\begin{\eq}
2 \varphi_n  \left(  \alpha r^2_{H}  +  \gamma k  -
      \gamma r_H  f_1\right) \left( q^2 A_t^2|_{r_H}+n^2 f_1  N^2_{1} \right)  +{\cal O}(\de)=0. \label{Taylor_Eq}
\end{\eq}
Since we are assuming $A_t|_{r_H} \neq 0$ and $\varphi_l \neq 0$ from the smoothness of the spaceime geometry, there are two ways to satisfy (\ref{Taylor_Eq}), at the leading order: (i) $\left(  \alpha r^2_{H}  +  \gamma k  -\gamma r_H  f_1\right)=0$, or (ii) $\left( q^2 A_t^2|_{r_H}+n^2 f_1  N^2_{1} \right)=0$.

Let us first assume that the case (i) is true and then, for $\ga \neq 0$, one can find
\begin{\eq}
f_1=\f{k}{r_{H}} +\left( \f{\al}{\ga} \right) r_H.
\label{Taylor_Eq_3}
\end{\eq}
But, the function $f_1$ in (\ref{Taylor_Eq_3}), which corresponds to the surface gravity $\ka_H \sim f'|_{r_H} \sim f_l$, is not generically what it should be:
$f'|_{r_H}$ has a wrong behavior either at the outer horizon $r_+$ or inner horizon $r_-$, where $f'|_{r_+}>0$ or $f'|_{r_-}<0$ are needed, respectively. Only for the special case of $k=-1, \al/\ga>0$, we may have the right behaviors of $f'|_{r_H}$. But, even in this also, (\ref{Taylor_Eq_3}) is problematic because its extremal point $r^2_H=-k \ga /\al$ where $f'|_{r_H}=0$, which corresponds to the extremal black holes with the vanishing Hawking temperature $T_H =\hbar \ka_H /2 \pi $, is just the function of theory parameters, $\al, \ga$, without the dependence of conserved quantities, like black hole mass and charges. Actually, this later property which is implied by (\ref{Taylor_Eq_3}) is a crucial to exclude the case of (i). On the other, for $\ga=0$, which is the case in Einstein-Maxwell-scalar theories \cite{Cai:2020}, the case (i) is trivially excluded. So, we find that the case (i) is not the correct choice.

Let us now assume that the case (ii) is true and then similarly one can find
\begin{\eq}
q^2 A_t^2|_{r_H}=-n^2 f_1  N^2_{1}.
\label{case_ii}
\end{\eq}
But this is impossible from two reasons. First, it is because the left hand side is positive, while the right hand side is {\it non-positive} (zero for the extremal horizons and positive otherwise), from the property of $f_1 \sim N^2_1$, up to a {\it positive} factor, for {\it Killing} horizons. This is basically the same argument as in \cite{Cai:2020}. Second, there is a more fundamental
and stronger argument against the above equation (\ref{case_ii}): No `$q$'
(charge of matter) appears in the right hand side, which all depend on the
black hole quantities (black hole mass, charges,  {\it etc}). Moreover,
$A_t|_{r_H}$ does not contain $q$ either but just depends on the black hole quantities
only: There is no way to compensate $q$ in the left hand side. In other words, the case (ii) is impossible unless $q=0$, or $A_t=0$ in contradiction to the assumption of case (b)! Note that this second argument is more fundamental and stronger than that of \cite{Cai:2020} against the choice of the case (b), without using the  property of Killing horizons!

So, we finally arrive that our assumption of the choice (b) can not be correct and then only the case (a) is the correct choice. Note that for the {\it neutral} scalar field, {\it i.e.}, $q=0$, one can not get any constraint on $A_t|_{r_H}$. All these proof are unchanged for any higher value of $l \geq n$ and arbitrary higher dimensions, where (\ref{E_varphi_Full}) becomes
\begin{\eq}
 &&r^{D-4}_{H}  \varphi_n  \left[  2 \alpha r^2_{H}  + (D^2-5 D+6) \gamma k  -(D-2)\gamma r_H  l f_l \de^{l-1} \right] \left( q^2 A_t^2|_{r_H} +n(l+n-1) f_l N^2_l \de^{2l-2}  \right) \no \\
  &&~~~~~~~~~~~+{\cal O}(\de^n)=0. \label{Taylor_Eq_Full}
\end{\eq}

\section{Lower-Dimensional Dilaton Gravity from Dimension Reduction}
{\label{app2}}
Here, we present the $(D-1)$-dimensional dilaton gravity from the dimensional reduction of $D$-dimensional EMH theory. Let us consider the {\it Kaluza-Klein} reduction of the $D$-dimensional metric $g_{\mu \nu}$ as
\begin{\eq}
ds^2=g_{\mu \nu} d x^{\m \n} \equiv e^{2 \sigma} \left(dz^2+h_{ab} dx^a dx^b \right),
\label{KK}
\end{\eq}
where the dilaton field $\sigma$ and metric $h_{ab}$ depend only on $(D-1)$-dimensional coordinates $x^a~ (a=0,1,2, \cdots, (D-1))$. In a straightforward computation \footnote{We thank Kyung Kiu Kim for sharing his unpublished work about this subject.}, one can find that
\begin{\eq}
R^{ab}&=&e^{-4 \si} \left\{
 \tilde{R}^{ab} -h^{ab} \left( \tilde{\nabla}^2 \si+(D-2) (\tilde{\nabla} \si)^2 \right)
+ (D-2)\left( \tilde{\nabla}^a \si \tilde{\nabla}^b \si - \tilde{\nabla}^a \tilde{\nabla}^b \si  \right)
\right\}, \\
R&=&e^{-2 \si} \left\{
 \tilde{R} -(D-2)(D-1) (\tilde{\nabla} \si)^2 -2(D-1) \tilde{\nabla}^2 \si
\right\}, \\
G^{ab}&\equiv& R^{ab}-\f{1}{2}  g^{ab} R \no \\
&=&e^{-4 \si} \left\{
 \tilde{R}^{ab}-\f{1}{2}  h^{ab} \tilde{R}
+(D-2)\left[ \tilde{\nabla}^a \si \tilde{\nabla}^b \si - \tilde{\nabla}^a \tilde{\nabla}^b \si +h^{ab} \left(\tilde{\nabla}^2 \si+ \f{(D-3)}{2} (\tilde{\nabla} \si)^2  \right)\right]
\right\}, \no \\
\end{\eq}
where $R=R^{ab} g_{ab}=e^{2 \si} R^{ab} h_{ab}$ and $\tilde{\nabla}^a, \tilde{R}^{ab}, \tilde{R}$ denote the covariant derivatives, Ricci tensor and scalar with respects to metric $h_{ab}$. Then, for the case of
\begin{\eq}
A=A_a(x^b) dx^a,
\label{KK_A}
\end{\eq}
{\it i.e.}, the absence of gauge potential $A_z$ along the compactification direction $z$, one can easily find that $D$-dimensional EMH theory (\ref{action}) reduces to, up to boundary terms,
\begin{\eq}
I_{(D-1)}&&=2 \pi \int dx^{D-1} \sqrt{-h}~ e^{(D-2) \si} \left\{ \tilde{R}+(D-2)(D-1) (\tilde{\nabla} \si)^2
-\f{Z(|\varphi|^2)}{4 \kappa} e^{-2 \si} \tilde{F}_{ab} \tilde{F}^{ab} \right. \no \\
&&-\al h^{ab}(\tilde{D}_{a} \varphi)^* (\tilde{D}_{b} \varphi)
 +\ga e^{-2 \si} \left[\tilde{R}^{ab}-\f{1}{2}  h^{ab} \tilde{R}+ (D-2) \left( \tilde{\nabla}^2 \si+\f{(D-3)}{2}(\tilde{\nabla} \si)^2  \right)
| \tilde{D} \varphi |^2  \no \right.\\
&& \left. \left. +(D-2) \left( \tilde{\nabla}^{a} \si \tilde{\nabla}^{b} \si -\tilde{\nabla}^{a} \tilde{\nabla}^{b} \si \right)(\tilde{D}_{a} \varphi)^* (\tilde{D}_{b} \varphi)\right] - e^{2 \si}V(|\varphi|^2)
 \right\},
\end{\eq}
where $\tilde{F}_{ab} \equiv \tilde{\nabla}_a A_b-\tilde{\nabla}_b A_a,~\tilde{D}_{a} \equiv \tilde{\nabla}_a-iq A_a$ and set the length scale of compact dimension $\int dz  \equiv 2 \pi$.
The reduced action corresponds the $(D-1)$-dimensional dilaton gravity with Maxwell and Horndeski couplings. By comparing (\ref{KK}), (\ref{KK_A}) with (\ref{Ansatz}), one finds that our spherically symmetric system corresponds to $e^{\si} \equiv r, h_{tt}\equiv-N^2/r^2, h_{rr} \equiv 1/f r^2$ with an identification of the azimuthal angular coordinate $\theta  \equiv z$ in $D$ dimensions. Note that the relevant conserved charge of $(D-1)$-dimensional dilaton gravity can be also obtained from the dimension reduction of the $D$-dimensional ${\cal Q}$ in (\ref{Q}). This indicates the validity of ``No Scalar-Haired Cauchy Horizon Theorem" for the dimensionally-reduced dilaton gravity also.

In particular, from the dimensional reduction of $D=3$ case, the $D=2$ dilaton gravity with Maxwell and Horndeski couplings is given by
\begin{\eq}
I_{(2)}&=&2 \pi \int dx^2 \sqrt{-h}~ e^{\si} \left\{ \tilde{R}+2 (\tilde{\nabla} \si)^2
-\f{Z(|\varphi|^2)}{4 \kappa} e^{-2 \si} \tilde{F}_{ab} \tilde{F}^{ab}
-\al h^{ab}(\tilde{D}_{a} \varphi)^* (\tilde{D}_{b} \varphi) \right. \no \\
&& \left. +\ga e^{-2 \si} \left[\tilde{\nabla}^2 \si | \tilde{D} \varphi |^2 +\left(\tilde{\nabla}^{a} \si \tilde{\nabla}^{b} \si -\tilde{\nabla}^{a} \tilde{\nabla}^{b} \si\right) (\tilde{D}_{a} \varphi)^* (\tilde{D}_{b} \varphi)\right] - e^{2 \si}V(|\varphi|^2)
 \right\},
 \label{D=2action}
\end{\eq}
where we have used the identity $\tilde{G}^{ab}\equiv \tilde{R}^{ab}-\f{1}{2}  h^{ab} \tilde{R} = 0$
in two dimensions. By identifying the $\si=\vp$, {\it i.e.}, a neutral scalar field $\vp$,
(\ref{D=2action}) reduces further to
\begin{\eq}
I_{(2)}&=&2 \pi \int dx^2 \sqrt{-h}~ e^{\si} \left\{ \tilde{R}+(2-\al) (\tilde{\nabla} \si)^2
-\f{Z(\si^2)}{4 \kappa} e^{-2 \si} \tilde{F}_{ab} \tilde{F}^{ab} \right. \no \\
&& \left. +\ga e^{-2 \si} \left[\tilde{\nabla}^2 \si (\tilde{\nabla} \si )^2 +\left(\tilde{\nabla}^{a} \si \tilde{\nabla}^{b} \si -\tilde{\nabla}^{a} \tilde{\nabla}^{b} \si\right) (\tilde{\nabla}_{a} \si) (\tilde{\nabla}_{b} \si)\right] - e^{2 \si}V(\si^2)
 \right\},
 \label{D=2action}
\end{\eq}
which covers a very general class of $D=2$ dilaton gravities \cite{Bank}.

Similarly, from the dimensional reduction of $D=4$ case, $D=3$ dilaton gravity with Maxwell and Horndeski couplings is given by
\begin{\eq}
I_{(3)}&&=2 \pi \int dx^{3} \sqrt{-h}~ e^{2 \si} \left\{ \tilde{R} +6 (\tilde{\nabla} \si)^2
-\f{Z(|\varphi|^2)}{4 \kappa} e^{-2 \si} \tilde{F}_{ab} \tilde{F}^{ab}
-\al h^{ab}(\tilde{D}_{a} \varphi)^* (\tilde{D}_{b} \varphi)  \right. \no \\
&&  +\ga e^{-2 \si} \left[\tilde{R}^{ab}-\f{1}{2}  h^{ab} \tilde{R}+ 2 \left( \tilde{\nabla}^2 \si+\f{1}{2}(\tilde{\nabla} \si)^2  \right)
| \tilde{D} \varphi |^2   +2 \left( \tilde{\nabla}^{a} \si \tilde{\nabla}^{b} \si -\tilde{\nabla}^{a} \tilde{\nabla}^{b} \si \right)(\tilde{D}_{a} \varphi)^* (\tilde{D}_{b} \varphi)\right] \no \\
&&\left. - e^{2 \si}V(|\varphi|^2)
 \right\}.
\end{\eq}

\newcommand{\J}[4]{#1 {\bf #2} #3 (#4)}
\newcommand{\andJ}[3]{{\bf #1} (#2) #3}
\newcommand{\AP}{Ann. Phys. (N.Y.)}
\newcommand{\MPL}{Mod. Phys. Lett.}
\newcommand{\NP}{Nucl. Phys.}
\newcommand{\PL}{Phys. Lett.}
\newcommand{\PR}{Phys. Rev. D}
\newcommand{\PRL}{Phys. Rev. Lett.}
\newcommand{\PTP}{Prog. Theor. Phys.}
\newcommand{\hep}[1]{ hep-th/{#1}}
\newcommand{\hepp}[1]{ hep-ph/{#1}}
\newcommand{\hepg}[1]{ gr-qc/{#1}}
\newcommand{\bi}{ \bibitem}

\end{document}